\documentstyle[11pt,newpasp,twoside,epsf]{article}
\reversemarginpar

\begin{document}
\title{Neutrino Properties from Measurements using Astrophysical and Terrestrial Sources}
 \author{A.B. McDonald}
\affil{Physics Department, Queen's University, Kingston, Ontario, Canada, K7L3N6}

\begin{abstract}
The current knowledge of neutrino properties has been derived from measurements performed
 with both astrophysical and terrestrial sources. Observations of neutrino flavor change
 have been made with neutrinos generated in the solar core, through cosmic ray interactions
 in the atmosphere and in nuclear reactors. A summary is presented of the current knowledge
 of neutrino properties and a description is provided for future measurements that could 
 provide more complete information on neutrino properties. 
\end{abstract}

\section{Introduction}

The neutrino has been very elusive in revealing its basic properties to experimenters. However,
 it provides a very attractive means for the study of many astrophysical objects such as the Sun, supernovae and 
other astrophysical sources producing high energy particles. The study of neutrinos from these sources can provide
information on both the sources and on basic properties of neutrinos themselves. This paper will discuss the
current state of information on neutrino properties, in several cases obtained from measurements with astrophysical
sources. Future neutrino measurements will be described for astrophysical or terrestrial sources. Other papers in this session will
 discuss measurements of astrophysical sources using this basic information on neutrino properties.

\section{Neutrino Properties}
\subsection{Number of neutrino types}

The number of active neutrino types has been restricted for many years by studies of Big Bang
 Nucleosynthesis (for a summary see Hagiwara et al, 2002) to be less than about 4, but considerably more accuracy
 has been obtained through measurements of the width of the $Z^0$ resonance that set a number of $2.994 \pm 0.012$. 
Neutrino flavor change measurements show no evidence for sterile neutrinos.

\subsection{Neutrino Flavor Change}

Several measurements have indicated that neutrino flavor change occurs; the most favored explanation for the
 mechanism is neutrino oscillations among finite mass eigenstates. The neutrino flavor
fields $\nu_{{\ell}}$ can be expressed as superpositions of the components
$\nu_{k}$ of the fields of neutrinos with definite masses $m_k$
via $U$, the 3 x 3 unitary Maki-Nakagawa-Sakata-Pontecorvo (MNSP) mixing matrix
(Maki et al, 1962, Gribov and Pontecorvo, 1969). 

The MNSP matrix can be parameterized
 in terms of 3 Euler angle rotations and a CP-violating phase $\delta_{CP}$,
$$ {\bf U} =
\left(\begin{array}{ccc}
c_{12}c_{13} & s_{12}c_{13} & s_{13}e^{-i\delta_{CP}} \\
-s_{12}c_{23} - c_{12}s_{23}s_{13}e^{i\delta_{CP}} &
c_{12}c_{23}-s_{12}s_{23}s_{13}e^{i\delta_{CP}} & s_{23}c_{13} \\
 s_{12}s_{23}-c_{12}c_{23}s_{13}e^{i\delta_{CP}} &
-c_{12}s_{23}-s_{12}c_{23}s_{13}e^{i\delta_{CP}} & c_{23}c_{13}
\end{array}
\right) 
$$
where $c_{ij}=\cos \theta_{ij} $ and $s_{ij}=\sin \theta_{ij} $.  

When the neutrinos travel in a vacuum or low density region and two mass eigenstates dominate the
 process, the following probability is predicted for subsequent detection of a given
 neutrino type after it has traveled for a distance L in vacuum:
$P = 1 - 1/2 sin^2 (2 \theta_{ij}) (1 - cos (2.54 \Delta m^2 L/E)$,
 where $\Delta m^2$ is the difference between the two
 relevant mass eigenstates in $eV^{2}$, L is the source-detector distance in meters, E is
 the neutrino energy in MeV and $\theta_{ij}$ is defined above.
 When the neutrinos pass through regions of high electron density, the difference in the interaction
 of electron neutrinos and other neutrinos due to the charged current interaction can add extra
 terms to the MNSP matrix, resulting in a change to the effective masses and coupling constants.
 This is referred to as the MSW effect (Mikeyev and Smirnov, 1985, Wolfenstein, 1978)
 and can be used to determine the sign of the mass difference of the
two dominant neutrinos involved in the oscillation.

\subsubsection{Atmospheric Neutrinos}
 
Super-Kamiokande has observed
 a zenith angle dependence that is consistent with flavor change of atmospheric muon neutrinos through
 oscillations with a baseline of the Earth's dameter. The zenith
 angle dependence for electron neutrinos is consistent with Monte Carlo calculations for no flavor change, implying
 that the muon flavor change is predominantly to tau neutrinos. The hypothesis of neutrino oscillations is
 consistent with measurements made by a number of other detectors of an anomalous ratio of muon to electron
 atmospheric neutrinos.

\subsubsection{Solar Neutrinos}

Since Davis' experiments starting in the 1960's, a discrepancy was identified
 between the experimental measurements and the theoretical calculations for solar neutrino fluxes.
The fluxes are factors of two or three lower than predictions
 in each case, leading to the conclusion that either solar models are incomplete or 
there are processes occurring such as flavor change to other neutrino types for which
 the experiments have little or no sensitivity. This 30-year old discrepancy had come
to be known as the "Solar Neutrino Problem".

Many attempts have been made to understand these discrepancies in terms of modifications to the solar
model, without significant success. The results may be understood in terms of neutrino flavor change with
 matter enhancement in the sun.  However, because the various experiments have different
thresholds and are sensitive to different combinations of the nuclear reactions in the sun, this explanation
 is solar model-dependent.
Solar model-independent approaches, including searches for spectral
distortion, day-night and seasonal flux differences provided no clear indication of flavor change. 

Measurements by the Sudbury Neutrino Observatory (SNO) of interactions 
of $^{8}B$ solar neutrinos in a heavy water detector have provided a solar-model-independent "appearance" measurement
 of neutrino flavor change by comparing charged current (CC) interactions on deuterium sensitive
 only to electron neutrinos and neutral current (NC) interactions sensitive to all neutrino types. A null hypothesis
 test for flavor change was performed, assuming no spectral change for the CC reaction.
The flux of active neutrinos or anti-neutrinos other than electron neutrinos inferred from
the NC measurements yielded a $5.3\sigma$ difference from the CC flux,
providing clear evidence for flavor change. The result for the total active neutrino flux obtained with the NC reaction,
$5.09^{+0.44}_{-0.43}(stat.) ^{+0.46}_{-0.43}(syst.)$,
is in very good agreement with the value calculated (Bahcall et al, 2001)
by solar models: $5.05\pm 1.0 \times 10^6~{\rm cm}^{-2} {\rm s}^{-1}$.

The solar neutrino measurements to date are best fit by neutrino oscillation parameters (see Table 1) 
including the MSW effect in the sun, referred to as the Large Mixing Angle (LMA) region.
 Note that the matter interaction defines $m_2$ to be greater
 than $m_1$ and that the mixing angle is somewhat smaller than maximal mixing.
 
\subsubsection{Terrestrial Measurements}
Measurements of the survival of $\nu_{\mu}$ neutrinos produced at the KEK accelerator have been made with the
 Super-Kamiokande detector, (K2K experiment). The preliminary data (Nishikawa, 2002) show agreement with the $m_2$ to $m_3$ oscillation
 parameters observed for atmospheric neutrinos. The KamLAND experiment (Eguchi et al, 2003) has studied the flux of electron anti-neutrinos 
observed at a 1000 ton liquid scintillator detector (converted from the original water-based Kamiokande detector).
 They find a
flux suppression consistent only with the LMA region for $m_1$ - $m_2$ oscillation as defined by solar neutrinos and
restricting the region obtained with solar neutrino measurements alone. Results from the LSND experiment
have indicated the appearance of a small flux of anti-$\nu_e$ from an anti-$\nu_\mu$ accelerator beam. The majority of the allowed
 oscillation region for this experiment has been restricted by the Karmen experiment. The MINIBOONE experiment has
 just begun operation with neutrino beams from Fermilab and should approach the LSND measurements with substantially
 higher sensitivity.

\subsubsection{Summary of flavor change information to date}

Atmospheric, solar, and reactor neutrino oscillation data currently fix or limit the 3 angles.  They also provide values
for the differences between the squares of the masses.  They provide no information yet on the phase(s). 
The data to date is summarized in Table 1.

\begin{table}[ht]
\caption{Current knowledge of active neutrino mass and mixing from neutrino oscillations. One-$\sigma$ errors are shown,
 except for $\theta_{13}$, which is at the 90\% CL.}
\label{angles}
\medskip 
\begin{center}
\begin{tabular}{lc}
\hline\hline
Quantity &  Value \\
\hline
$\theta_{12}$ & 32.6(32)$^o$ \\
$\theta_{13}$ & $<10^o$ \\
$\theta_{23}$ & 45(8)$^o$ \\
$\delta_{CP}$ & ? \\
$m_2^2-m_1^2$ & $+ 7.3(11) \times 10^{-5}$  eV$^2$  \\
$m_3^2-m_2^2$ & $\pm  2.5(6) \times 10^{-3}$  eV$^2$ \\
\hline
\end{tabular}
\end{center}
\end{table}

\subsection{Neutrino Mass}

The most sensitive direct measurements of electron neutrino mass have been made by searching for curvature induced
near the end point of the spectrum of electrons emitted during the beta decay of tritium. The current limit obtained
 from these measurements is 2.8 eV (90 \% CL.). Measurements of neutrinoless double beta decay are also sensitive to
 neutrino mass if the neutrino is a Majorana particle. Measurements to date set limits less than 0.4 eV for the
 effective mass associated with this process. There is also a controversial claim of a greater than 2 $\sigma$  effect
for a mass of 0.35 eV in a neutrinoless double beta decay measurement 
in Ge reported by a subset of the Heidelberg-Moscow experimental group. Model-dependent limits
 with sensitivity of about 1 eV can also be obtained
 from combined fits to the cosmic microwave and large scale structure data.

\section{Future measurements}

All of the types of measurements discussed above are being pursued very actively for the future. The next generation
 measurements for tritium beta decay and neutrinoless double beta decay should extend the mass sensitivity by a factor
 of about 10. This sensitivity is approaching the mass differences identified by the oscillation measurements. Flavor
 change measurements will be extended for solar and terrestrial neutrinos with improved accuracy. The definition
 of these parameters has also led to plans for a next generation of 
long-baseline experiments to quantify $\theta_{13}$ through accelerator and reactor experiments and seek the mass
 hierarchy through matter interactions and CP
 violating phase through experiments with accelerator and detector properties scaled up by factors of 10.

Our knowledge of neutrino properties has expanded greatly during the past 10 years. The next generations of experiments
 have the potential to provie as comprehensive information for the lepton sector as has been obtained for quarks, thereby
making it possible to use neutrinos as a unique astrophysical probe.

\end{document}